\documentclass[letterpaper, 10 pt, conference]{ieeeconf}  %

\IEEEoverridecommandlockouts                              %

\newcommand{\etal}{\textit{et al.}}
\usepackage{graphicx}
\usepackage{xcolor}
\usepackage{soul}
\graphicspath{{images/}}
\usepackage{tensor}
\usepackage{subfig}

\usepackage{mathtools}
\usepackage{algorithm}
\usepackage[noend]{algpseudocode}
\algrenewcommand\algorithmicrequire{\textbf{Input:}}
\algrenewcommand\algorithmicensure{\textbf{Output:}}
\newcommand*\Let[2]{\State #1 $\gets$ #2}
\algnewcommand\algorithmicforeach{\textbf{for each}}
\algdef{S}[FOR]{ForEach}[1]{\algorithmicforeach\ #1\ \algorithmicdo}
\DeclarePairedDelimiterX{\norm}[1]{\lVert}{\rVert}{#1}

\usepackage{outlines}
\usepackage{booktabs}
\usepackage[pagebackref=true,breaklinks=true,letterpaper=true,colorlinks,bookmarks=false]{hyperref} %

\title{\LARGE \bf
Online Photometric Calibration of \\Automatic Gain Thermal Infrared Cameras
}

\author{Manash Pratim Das$^{1, 2}$, Larry Matthies$^{1}$ and Shreyansh Daftry$^{1}$%
\thanks{$^{1}$Jet Propulsion Laboratory, California Institute of Technology, CA, USA, {\tt\small \{lhm,daftry\}@jpl.nasa.gov}. $^{2}$Carnegie Mellon University, PA, USA {\tt\small mpdmanash@cmu.edu}. Copyright \copyright  2020. California Institute of Technology. U.S. Government sponsorship acknowledged.}%
}

\usepackage{fancyhdr}
\begin{document}

\maketitle
\pagestyle{empty}

\begin{abstract}

Thermal infrared cameras are increasingly being used in various applications such as robot vision, industrial inspection and medical imaging, thanks to their improved resolution and portability. However, the performance of traditional computer vision techniques developed for electro-optical imagery does not directly translate to the thermal domain due to two major reasons: these algorithms require photometric assumptions to hold, and methods for photometric calibration of RGB cameras cannot be applied to thermal-infrared cameras due to difference in data acquisition and sensor phenomenology. In this paper, we take a step in this direction, and introduce a novel algorithm for online photometric calibration of thermal-infrared cameras. Our proposed method does not require any specific driver/hardware support and hence can be applied to any commercial off-the-shelf thermal IR camera. We present this in the context of visual odometry and SLAM algorithms, and demonstrate the efficacy of our proposed system through extensive experiments for both standard benchmark datasets, and real-world field tests with a thermal-infrared camera in natural outdoor environments. 

\end{abstract}

\section{Introduction}
\label{sec:intro}
In the last decade, tremendous efforts have been devoted towards the development of computer vision methods to autonomous robotic systems. Due to their success, a broad range of critical applications from disaster scene surveillance to autonomous driving and package delivery have been made possible already, or are well within reach. However, most of these approaches have addressed only daytime operation under normal illumination, using standard visible-spectrum cameras. For applications that are not resource-constrained and allow the use of active sensors such as LiDARs for night-time operation, they still do not scale to other visually-degraded environmental effects such as fog, smoke, and dust \cite{brunner2014perception}. This is very limiting, and enabling operations to more challenging conditions would significantly enhance the tactical value of such systems. 

In this paper, we argue that thermal infrared (TIR) cameras offer a compelling and complementary advantage to both visible-spectrum cameras and LiDARs. The nature of the TIR modality makes them highly robust to both low-illumination conditions and in the presence of adverse obscurants \cite{beier2004simulation}. Furthermore, due to the rapid growth in sensory technology \cite{jeong2018development}, the quality of images and frame-rate generation of low SWaP, COTS, uncooled micro-bolometer arrays ($\mu$BA) based TIR cameras have increased to a level that is suitable for common computer vision tasks. As a result, there has been a growing interest in the computer vision community to use thermal cameras - ranging from object detection \cite{terryobject} and activity recognition \cite{han2005human} to stereo \cite{treible2017cats}, visual odometry \cite{mouats2015thermal,borges2016practical}, SLAM \cite{shin2019sparse,vidas2012hand} and 3D mapping \cite{daftry2018robust}. See \cite{gade2014thermal} for a comprehensive review on computer vision research and applications with thermal infrared cameras.

\begin{figure*}[t!]
	\centering
	\subfloat{\includegraphics[width=0.16\textwidth]{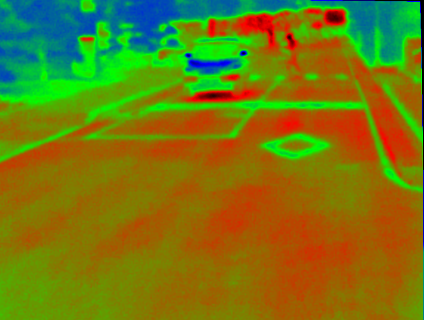}}
	\hfill
	\subfloat{\includegraphics[width=0.16\textwidth]{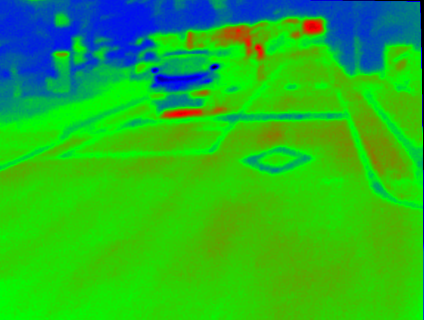}}
	\hfill
	\subfloat{\includegraphics[width=0.16\textwidth]{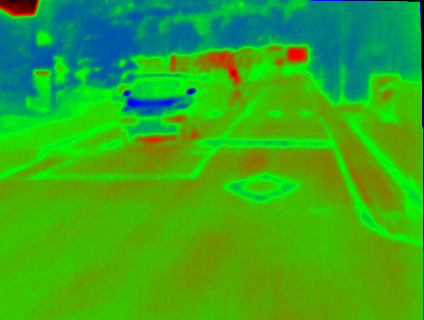}}
	\hfill
	\subfloat{\includegraphics[width=0.16\textwidth]{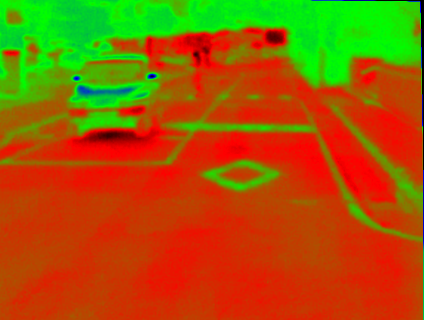}}
	\hfill
	\subfloat{\includegraphics[width=0.16\textwidth]{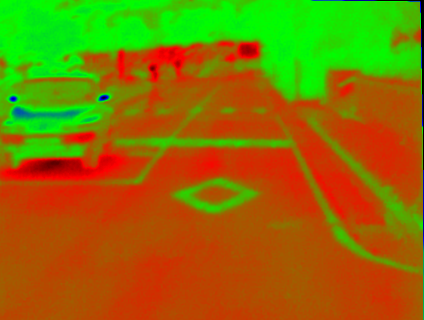}}
	\hfill
	\subfloat{\includegraphics[width=0.16\textwidth]{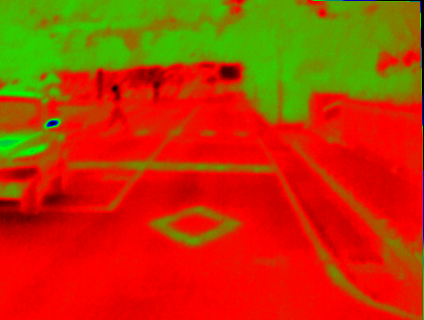}}	
	\addtocounter{subfigure}{-6}
	\subfloat{\includegraphics[width=0.16\textwidth]{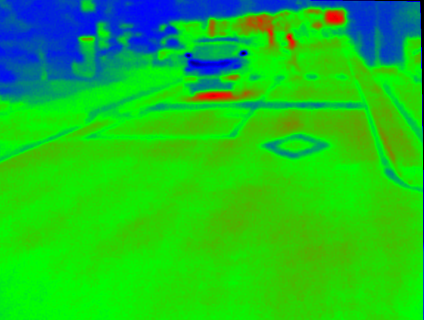}}
	\hfill
	\subfloat{\includegraphics[width=0.16\textwidth]{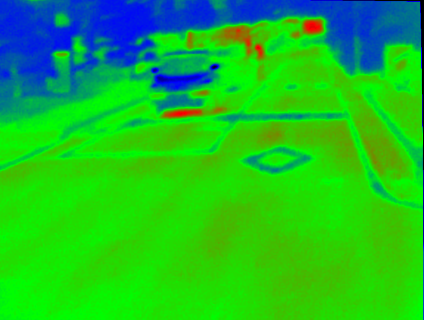}}
	\hfill
	\subfloat{\includegraphics[width=0.16\textwidth]{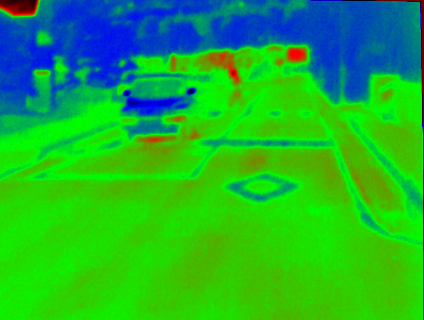}}
	\hfill
	\subfloat{\includegraphics[width=0.16\textwidth]{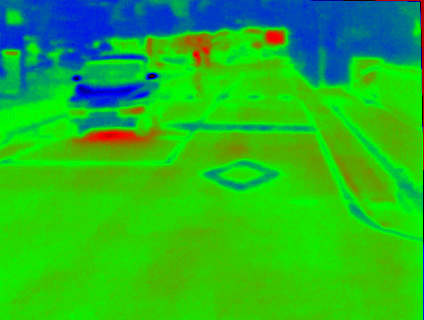}}
	\hfill
	\subfloat{\includegraphics[width=0.16\textwidth]{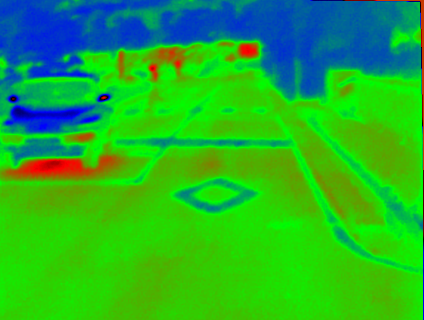}}
	\hfill
	\subfloat{\includegraphics[width=0.16\textwidth]{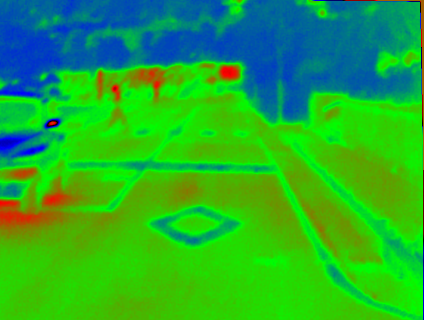}}
	
	\caption{\small In this paper we propose an approach for full photometric calibration of automatic gain thermal-infrared videos, which can be used either offline for calibrating existing datasets, or online in combination with state-of-the-art VO or SLAM pipelines. \textbf{Top:} Sequence of thermal images from the KAIST dataset \cite{choi2015all} with strong exposure changes. \textbf{Bottom:} the same sequence after photometric calibration.}
	\label{fig:qualitative}
	\vspace{-0.2cm}
\end{figure*}

Despite the aforementioned advantages, the performance of traditional computer vision techniques developed for electro-optical imagery does not directly translate to the thermal domain \cite{daftry2018robust,khattak2019keyframejfr}. This is primarily because these cameras have several characteristics that are challenging for vision algorithms. In particular, for visual odometry and SLAM algorithms, one of the biggest challenges is photometric inconsistency due to rapid automatic gain change (AGC). As hot objects move in and out of the camera’s field of view, large regions of pixels saturate, and the sensor adjusts gain to darken the image in an attempt to avoid saturation. The resulting change in intensity can be dramatic from one frame to the next (as shown in \autoref{fig:qualitative}), invalidating the constant brightness assumption \cite{shin2019sparse} enforced in most low-level computer vision tasks (both direct or feature-based). 

It is be noted that the actual measured radiometric values of the objects in the environment are not affected by the observation of other objects. On the contrary, only the appearance of objects in the rescaled thermal images changes as a response to the rescaling operation applied to full radiometric data. One easy fix is to disable AGC completely \cite{papachristos2018thermal}, perform sensor-specific photometric calibration \cite{budzier2015calibration} and manually set a fixed thermal range for the entire operation. This requires prior knowledge of the particular environment, and this limits the generalization of methods and applicability to long-term navigation tasks \cite{chen2017rgb}. Furthermore, one might want to run a visual odometry or SLAM algorithm on datasets where no photometric calibration is provided and no access to the camera is given. In these cases, it is necessary to use an algorithm that can provide photometric calibrations for arbitrary video sequences. While such approaches exist for visual spectral cameras \cite{bergmann2018online,goldman2010vignette,kim2010joint}, the same photometric model cannot be used for TIR cameras due to different data acquisition mechanism, sensor phenomenology and non-uniform properties of the $\mu$BA.

In this paper, our goal was to develop an approach for full photometric calibration of automatic gain TIR cameras, which can be used either offline for calibrating existing datasets, or online in combination with state-of-the-art visual odometry or SLAM pipelines. Specifically, our contributions are threefold: First, we develop a new photometric model that accounts for 1) automatic temporal changes in gains, and 2) spatial biases in pixel intensities of TIR cameras. Second, we present an efficient multi-threaded algorithm that estimated the parameters of the proposed model in real-time by solving multiple non-linear least square problems. Finally, we demonstrate that our proposed calibration approach benefits various  downstream  computer  vision pipelines through extensive experiments with both standard datasets and real-world field tests with a TIR camera.

\section{Photometric Model for TIR Cameras}
\label{model}
\begin{figure}[t!]
	\subfloat{\label{fig:Microbolometer}\includegraphics[width=0.45\linewidth]{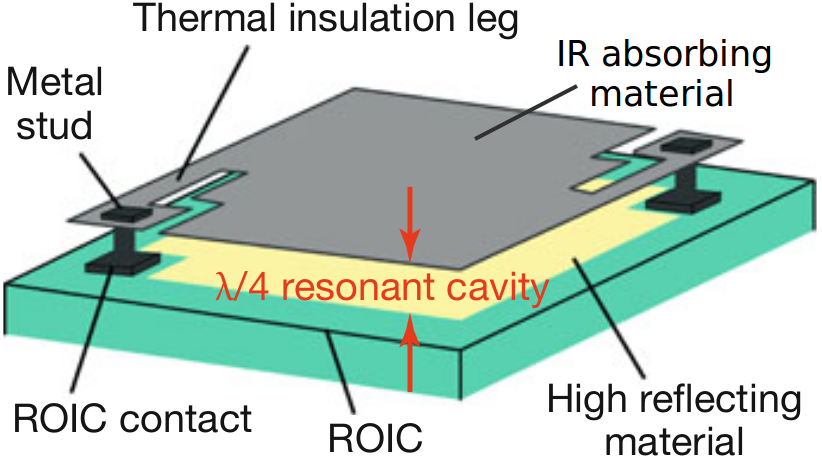}}
	\hfill%
	\subfloat{\label{fig:Emission}\includegraphics[width=0.4\linewidth]{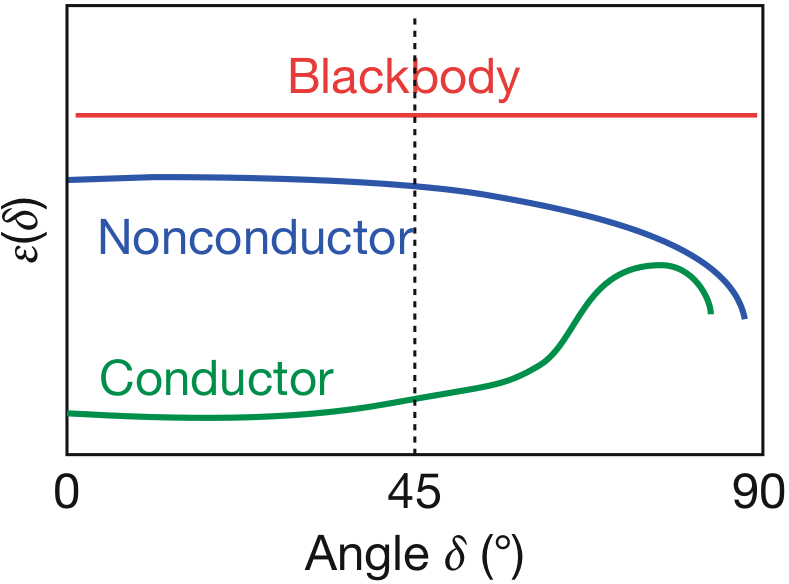}}
	\caption{\small \textbf{Left}: A generalized cross-sectional view of a $\mu$BA pixel. \textbf{Right}: Emissivity $\varepsilon(\delta)$ as a function of vieweing angle $\delta$ for blackbody, nonconductor and conductor. Source~\cite{vollmer2017infrared}}
\end{figure}

A microbolometer is an temperature-dependent electrical resistor (as illustrated in~\autoref{fig:Microbolometer}). Infrared radiation with wavelengths between $7.5 - 14 \mu m$ strikes the detector material, heating it, and thus changing its electrical resistance. This resistance change is measured by a readout integrated circuit (ROIC) and processed into temperatures. Many such microbolometers are arranged in a 2D grid to form a focal plane $\mu$BA, where temperature reading from each microbolometer is used to generate intensity for the corresponding pixel in the image.

The infrared energy radiated by an object comprises of the energy emitted, transmitted and reflected by the object. For our photometric model, we assume that the infrared emissivity of an object is invariant to the viewing angle of an observer. While this is true only for a blackbody object, Vollmer \etal ~\cite{vollmer2017infrared} upon studying a wide variety of materials observed that for nearly all practically relevant surfaces, the emissivity is nearly constant up to at least $\pm45^\circ$ viewing angle (as illustrated in~\autoref{fig:Emission}). Furthermore, outside this range, for a nonconductor dominant scene if our viewing angle does not change much, the emissivity would also not change much until extreme viewing angles $>78^\circ$.

Now consider a 3D world point in the scene which is emitting a constant amount of infrared energy with radiant power $\Phi_o$. The change in electrical resistance of the microbolometer is given by \autoref{eq:delr1}, where $\alpha \Phi_o$ is absorbed radiant power, $\beta$ is thermal coefficient and $G_\textnormal{th}$ is total thermal conductance. The change in resistance can also be expressed in terms of the change in temperature $\Delta T$ of the microbolometer as \autoref{eq:delr2} where $\overline{\beta R}$ is the average value in the temperature interval $\Delta T$. Refer to \cite{vollmer2017infrared} for more details on response generation.

\noindent  
\begin{minipage}{0.45\columnwidth}
	\begin{equation}
	\label{eq:delr1}
	\Delta R = \frac{\beta R \alpha \Phi_o}{G_\textnormal{th}}
	\end{equation}
\end{minipage}
\begin{minipage}{0.45\columnwidth}
	\begin{equation}
	\label{eq:delr2}
	\Delta R = \overline{\beta R}\Delta T
	\end{equation}
\end{minipage}
\vspace{2mm}

\noindent
Now. after simplification, if the microbolometer is sensitive to an energy range given by $[\Phi_\textnormal{min}, \Phi_\textnormal{max}]$, then the normalized response from the sensor is given by $I = (\Phi-\Phi_\textnormal{min}) / (\Phi_\textnormal{max}-\Phi_\textnormal{min})$. This response $I$ is used to generate intensity value for the corresponding pixel in the synthetic image $\mathcal{I} : \Omega \rightarrow [0,1]$. Here $\Omega$ denotes the image domain.

\begin{figure}[t!]
	\subfloat[]{\label{fig:rgb_spatial}\includegraphics[width=0.24\linewidth]{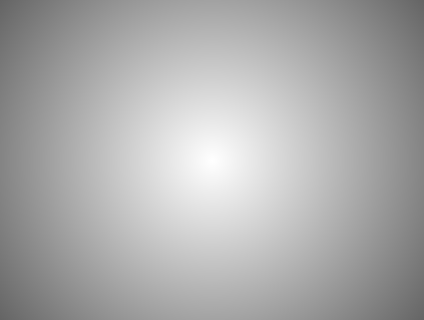}}\hfill%
	\subfloat[]{\label{fig:ir_spatial}\includegraphics[width=0.24\linewidth]{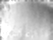}}
	\hfill%
	\hfill%
	\hfill%
	\subfloat[]{\label{fig:NUC}\includegraphics[width=0.24\linewidth]{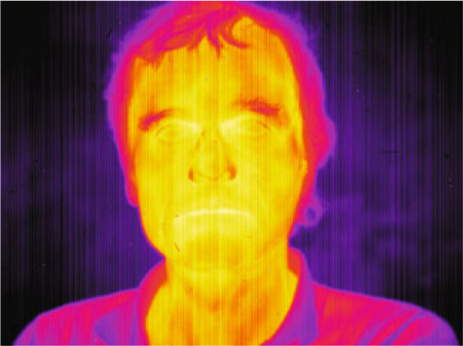}}\hfill%
	\subfloat[]{\includegraphics[width=0.24\linewidth]{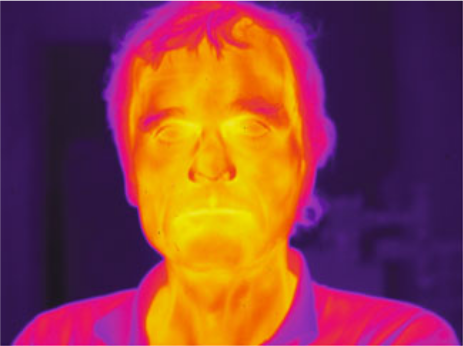}}
	\caption{\small Visualization of (a) radial lens vignetting used for visual spectral cameras \cite{bergmann2018online}, as compared to (b) non-symmetric spatial variation in $\mu$BA-based TIR camera estimated by our method. TIR images (c) before and (d) after NUC performed by the manufacturer on the hardware (note the streaks) (Source~\cite{vollmer2017infrared}).}
	\vspace{-5mm}
\end{figure}

We observe that the image intensity of a scene point with constant thermal radiant energy varies when viewed at a different spatial location in the image $x\in\Omega$. In visible spectrum cameras, this spatially varying image intensity is contributed to radial-lens vignetting (\autoref{fig:rgb_spatial}) and modelled as a sixth-order polynomial \cite{goldman2010vignette}, assuming that the attenuation factors follow a radial model and the center of vignetting coincides with the image center. In regards to TIR cameras, the spatial variation does not take a predefined or symmetric shape (\autoref{fig:ir_spatial}). The variation can be due to two major reasons: 1) the response of each microbolometer might be different either due to the design properties or saturation of the sensor that leads to residuals in their reading, and 2) any infrared transmission effects that lead to uneven transparency including lens effects. The infrared thermal imaging community is well aware of the non-uniform responses by the individual microbolometers in the array. Therefore the non-uniform correction (NUC) (\autoref{fig:NUC}) to correct for the different gains and signal offsets of individual detectors has been widely studied~\cite{orzanowski2016nonuniformity,bieszczad2009method,budzier2015calibration}. While most of the hardware will come pre-calibrated, we do not assume the same in our approach to generalize over wide range of cameras. For photometric calibration we are more concerned about low-frequency non-uniformness; variation that spreads across a large area of pixels, and is significant enough to introduce photometric errors. The microbolometers can also get heated by the camera device itself, most commonly due to uneven saturation during operation and the lack of cooling. Therefore the low cost uncooled  $\mu$BA-based cameras are more susceptible to low-frequency spatial variations. 

We model this effect using an affine sensor response given by $I'_x = s_x(I_x + r_x)$. However, we observe that using only the bias $r_x$ and ignoring the scale factor $s_x$ is sufficient to model this low-frequency effect while reducing the number of parameters by half. Thus, the spatial bias is modeled by an offset $r_x$ in the pixel is given by \autoref{eq:offset}. We also observe that the camera tries to maintain best possible temperature resolution at each time (or frame) $t$ in the image and therefore we can express $I'_x$ as a normalized value (\autoref{eq:gain}) within a range $[I'_{t,\textnormal{min}}, I'_{t,\textnormal{max}}]$, where $I'_{t,\textnormal{min}}$ and $I'_{t,\textnormal{max}}$ are the minimum and maximum responses respectively among all the microbolometers for the particular frame $t$.

\noindent
\begin{minipage}{0.35\columnwidth}
	\begin{equation}
	I'_x = I_x + r_x \label{eq:offset}
	\end{equation}
	\vspace{0.2mm}
\end{minipage}%
\begin{minipage}{0.55\columnwidth}
	\begin{equation}
	I'_{x,t} = \frac{I'_x-I'_{t,\textnormal{min}}} {I'_{t,\textnormal{max}} - I'_{t,\textnormal{min}}} \label{eq:gain}
	\end{equation}
\end{minipage}

\noindent Thus, the raw response from the sensor after spatial error is shifted by $I'_{t,\textnormal{min}}$ and scaled by $(I'_{t,\textnormal{max}}-I'_{t,\textnormal{min}})^{-1}$ before being used to generate the image. This is a major contributor to the effect commonly known as gain change or AGC. This implies that the scene point with constant thermal radiant energy may appear with different image intensities in consecutive frames.

Finally, combining \autoref{eq:offset} and \autoref{eq:gain}, the normalized temperature response is given by \autoref{eq:affine_model}. In \autoref{eq:log_affine_model} we describe the gain change as an affine brightness response function, similar to \cite{engel2018direct} for visual spectral images, where $e^{a_t} = I'_{t,\textnormal{max}} - I'_{t,\textnormal{min}}$ and $b_t = I'_{t,\textnormal{min}}$. 

\noindent
\begin{minipage}{0.5\columnwidth}
	\begin{equation}
	\label{eq:affine_model}
	I'_{x,t} = \frac{I_x + r_{x} -I'_{t,\textnormal{min}}} {I'_{t,\textnormal{max}} - I'_{t,\textnormal{min}}}
	\end{equation}
	\vspace{1mm}
\end{minipage}
\begin{minipage}{0.45\columnwidth}
	\begin{equation}
	\label{eq:log_affine_model}
	I'_{x,t} = \frac{I_x + r_{x} -b_t} {e^{a_t}}
	\end{equation}
	\vspace{2mm}
\end{minipage}

\noindent The logarithmic parameterization of the scale factor $e^{-a_t}$ prevents it from becoming negative and makes it numerically stable. The goal of the proposed photometric calibration then is to attempt to estimate $a_t$ and $b_t$ for each frame and $r_{x}$ for each pixel such that, a scene point in the world with constant radiant energy should appear with the same image intensity value invariant to temporal frames and spatial locations.

\section{Algorithmic Framework}
\label{sec:algo}
\subsection{Problem Formulation}
Let $p,m,n \in \Omega$ be the pixel locations for the same 3D world point at frames $o,t \text{ and } t+1$ respectively, thus, $I_p=I_m=I_n$. Without loss of generality, consider $o$ to be a hypothetical frame where $e^{a_o}=1$, $b_o=0$ and without any residual errors, such that $I'_{p,o}=I_p$. At other frames $t,t+1$, we can consider the parameters be represented with respect to the hypothetical frame $o$. Now using~\autoref{eq:log_affine_model}
\begin{align}
I'_{m,t} &= (I'_{p,o}-{}^ob_t+{}^pr_m ) / e^{{}^oa_t} \label{eq:calib_model} \\
I'_{n,t+1} &= (I'_{p,o}-{}^ob_{t+1}+{}^pr_n ) / e^{{}^oa_{t+1}} \nonumber \\
&= (I'_{m,t}e^{{}^oa_t} + {}^ob_t-{}^pr_m-{}^ob_{t+1}+{}^pr_n)/ e^{{}^oa_{t+1}} \nonumber \\
&= f_P(I'_{m,t})
\end{align}
where, $P$ represents the set of model parameters and ${}^pr_m=r_m-r_p$. We factor out the intensity $I'_{p,o}$ of the hypothetical frame as it is unobservable, and decompose the function $f_P(I'_{m,t})$ as follows:

\begin{align}
\label{eq:ghfunctions}
f_P(I'_{m,t}) &= \frac{I'_{m,t}e^{{}^oa_t} + {}^ob_t-{}^ob_{t+1}}{e^{{}^oa_{t+1}}}  + \frac{{}^pr_n-{}^pr_m}{e^{{}^oa_{t+1}}} \nonumber \\
&= g_{P_t}(I'_{m,t}) + h_{P_t,\mathcal{P}_s}
\end{align}
where $P = P_t \cup \mathcal{P}_s$ are the unknown set of parameters that relates to temporal and spatial variations respectively. Thus, there are two unknown parameters ($\{{}^oa_t,{}^ob_t\} \subset P_t$) per frame, and additional $\mathcal{P}_s = \mathcal{O}(| \Omega |)$ unknown parameters. Our goal is to estimate these parameters efficiently in a real-time online setting using pixel correspondences across frames (see \ref{sec:pixel_correspondences} for details regarding correspondences).

\subsection{Measurement Model}
\label{measurement-model}
Let $\epsilon$ be zero-mean Gaussian noise, we consider our measurement model as
\begin{align*}
I'_{n,t+1} &= g_{P_t}(I'_{m,t}) + h_{P_t,\mathcal{P}_s} + \epsilon \\
& =  g_{P_t}(I'_{m,t}) + \frac{ \Delta_{n,m} }{ e^{{}^oa_{t+1}} } + \epsilon
\end{align*}
where $\Delta_{n,m} = {}^pr_n-{}^pr_m$. If we ignore the $h_{P_t,\mathcal{P}_s}$ term, and consider the estimation of only the $P_t$ parameters, the Maximum Likelihood Estimate of $P_t$ corresponds to the solution of the least squares objective given by
\begin{align}
\begin{split}
\label{eq:tobjective}
O_\text{T}(g_{P_t}, C_t) =& \min\limits_{P_t} \sum_{(m,n)\in C_t} \norm{I'_{n,t+1}-g_{P_t}(I'_{m,t})}^2
\end{split}
\end{align}
where $C_t$ denotes the set of pixel correspondences between frame $t$ and $t+1$. Note that, the term $\Delta_{n,m} /  e^{{}^oa_{t+1}}$ becomes part of the error, however unlike $\epsilon$, it is not zero-mean. Thus, minimizing the objective function~\autoref{eq:tobjective} favors the minimization of $\Delta_{n,m} /  e^{{}^oa_{t+1}}$. This has an adverse effect on ${}^oa_{t+1}$, as when $\Delta_{n,m} > 0$, the term $e^{{}^oa_{t+1}}$ gets increased, and vice versa.
However, if we do not ignore the $h_{P_t,\mathcal{P}_s}$ term, on one hand, we can prevent the aforementioned bias, but on the other hand, our objective function in~\autoref{eq:tsobjective} includes the additional unknown terms $\mathcal{P}_s$
\begin{align}
\begin{split}
\label{eq:tsobjective}
O_\text{TS}(f_{P_t,\mathcal{P}_s}, C_t) =& \min\limits_{P_t,\mathcal{P}_s} \sum_{t=1}^{N} \sum_{(m,n)\in C_t} \norm{I'_{n,t+1}-f_{P_t,\mathcal{P}_s}(I'_{m,t})}^2
\end{split}
\end{align}

\subsection{Estimating Temporal Parameters}
\label{estimating-temporal}
There are two unknown temporal parameters (${}^oa_t, {}^ob_t$) per frame. We desire to update them with minimal computational overhead in order to maintain high frame rate throughput. Using either objective functions \autoref{eq:tobjective} or \autoref{eq:tsobjective}, we can iteratively solve for ${}^oa_t, {}^ob_t$ using ${}^oa_{t-1}, {}^ob_{t-1}$, and point correspondences between frame $(t-1)$ and $t$. If we use the latter, we need to either also estimate $\mathcal{P}_s$ or fix them with known values. Alternatively, if we use the former, we face the problem of bias as discussed in~\ref{measurement-model}. To solve this, we observe that spatial variations occur in small regions on the image, and thus, we assume that majority of the pixel correspondences have $\Delta_{n,m} \approx 0$, while those with $\Delta_{n,m} >> 0$ will form the minority. We employ RANSAC \cite{fischler1981random} to reject the outliers and estimate $P_t$.   

Next, for any pair ($t,t+1$) of frames in consideration, instead of estimating all the unknown parameters ($\{{}^oa_{t}, {}^ob_t, {}^oa_{t+1}, {}^ob_{t+1}\}$) w.r.t.\ the hypothetical frame $o$, we instead estimate the relative parameters ($\{{}^ta_{t+1}, {}^tb_{t+1}\}$)
\begin{minipage}{0.4\columnwidth}
	\begin{align}
	\label{eq:rela}
	e^{{}^ta_{t+1}} = \frac{e^{{}^oa_{t+1}}}{ e^{{}^oa_{t}} }
	\end{align}
\end{minipage}
\begin{minipage}{0.59\columnwidth}
	\begin{align}
	\label{eq:relb}
	{}^tb_{t+1} = \frac{{}^ob_{t+1} - {}^ob_{t}}{ e^{{}^oa_{t}} }
	\end{align}
\end{minipage}

\begin{align}
e^{{}^ta_{t+1}} &= e^{{}^ta_{t+1}} \times e^{{}^{t+1}a_{t+2}} \label{eq:chaina} \\    
{}^tb_{t+2} &= {}^tb_{t+1} + e^{{}^ta_{t+1}} \times {}^{t+1}b_{t+2} \label{eq:chainb}
\end{align}

\noindent
Thus, using \autoref{eq:ghfunctions}, \autoref{eq:rela}, \autoref{eq:relb} and ignoring $h_{P_t,\mathcal{P}_s}$
\begin{align}
\label{eq:g'functions}
I'_{n,t+1} = (I'_{m,t} - {}^tb_{t+1} ) / e^{{}^ta_{t+1}} = g'_{P'_t}(I'_{m,t})
\end{align}
Until now, we considered only consecutive frames ($t,t+1$) in our discussion to indicate a sense of sequential and progressive process. However, in the presence of pixel correspondences $C_{i,j}$, the above concepts would generalize to any two frames $i,j$, and  the relative parameters $P'_{i,j} = \{ {}^ia_j, {}^ib_j \}$ can be estimated using $O_\text{T}(g'_{P'_{i,j}}, C_{i,j})$ in \autoref{eq:tobjective}. In-fact a current frame $t$ can have pixel correspondences with multiple previous frames $t-1, t-2, t-k,\ldots$, and all of these correspondences are used to get the best estimate for $P'_{t-1,t} = \{ {}^{t-1}a_t, {}^{t-1}b_t \}$ as described in~\autoref{alg:temporal_algorithm}. Thus, we again overload the notation for $C_t = C_{t-1,t} \cup C_{t-2,t} \cup C_{t-k,t} \cup \ldots$ to denote the set that contains all the pixel correspondences from previous frames to the current frame $t$. Despite of having correspondences from multiple previous frames, the estimates ($P'_{t-1,t}$) might still incur drift error, hence, in \textsc{AdjustForDrift}, we attempt to keep the parameters close to their nominal values. Finally, let $P'_{t-1} = \bigcup^{t-1}_{x=t-N} P'_{1,x}$ be the set of all parameters for $N$ frames relative to the first frame, where $P'_{1,1} = \{{}^1a_1 = 1, {}^1b_1=0\}$, and for all other frames $x \neq 1$, $P'_{1,x}$ is computed by chaining all $P'_{y-1,y}, y = [2 \mathrel{{.}\,{.}}\nobreak x]$ using \autoref{eq:chaina} and \autoref{eq:chainb}. We employ multi-threading to run the iterations of the ``\textbf{for}'' loop in line 4 \autoref{alg:temporal_algorithm} as they are independent.

\begin{algorithm}[t!]
	\caption{Estimating temporal parameters
		\label{alg:temporal_algorithm}}
	\begin{algorithmic}[1]
		\Require{$C_t$, $P'_{t-1}$, $N$}
		\Ensure{$P'_{t}$, the set of all relative parameters upto frame $t$}
		
		\Function{Main}{$C_t,P'_{t-1},N$}
		\Let{$T$}{$0$}
		\Let{$\{{}^{t-1}a_t, {}^{t-1}b_t\}$}{$\{0,0\}$}
		\ForEach {$C_{i,t} \in C_t $ asynchronously in parallel}
		\Let{$P'_{i,t}$}{\textsc{Ransac} on \autoref{eq:tobjective} using $O_\text{T}(g'_{P'_{i,t}}, C_{i,t})$}
		\Let{$a^i,b^i$}{\textsc{ChangeRef}($P'_{i,t}$, $P'_{1,i}$, $P'_{1,t-1}$)}
		\Let{$ {}^{t-1}a_t $}{$ {}^{t-1}a_t + a^i \times |C_{i,t}|$}
		\Let{$ {}^{t-1}b_t $}{$ {}^{t-1}b_t + b^i \times |C_{i,t}|$}
		\Let{$T$}{$T+|C_{i,j}|$}
		\EndFor
		\Let{$\{ {}^{t-1}a_t, {}^{t-1}b_t \}$}{$\{ {}^{t-1}a_t/T, {}^{t-1}b_t/T \}$}
		\Let{$\{ {}^{t-1}a_t, {}^{t-1}b_t \}$}{\textsc{AdjustForDrift}($ {}^{t-1}a_t, {}^{t-1}b_t $)}
		
		\Let{$\{ {}^1a_t, {}^1b_t \}$}{$\{ {}^1a_{t-1} + {}^{t-1}a_t, {}^1b_i + e^{{}^1a_{t-1}} \times {}^{t-1}b_t \}$}
		\Let{$P'_{t}$}{$P'_{t-1} \cup \{\{{}^1a_t, {}^1b_t\}\}$}
		\State \Return {$P'_{t}$}
		\EndFunction

		\Function{ChangeRef}{$P'_{i,t},P'_{1,i}, P'_{1,t-1}$}
		\Let{$ \{ {}^ia_{t-1}, {}^ib_{t-1}\}$}{$\{ {}^1a_{t-1} - {}^1a_{i},  ({}^1b_{t-1}-{}^1b_{i}) / e^{{}^1a_{i}}  \}$}
		\Let{$\{ {}^{t-1}a_{t}, {}^{t-1}b_{t} \}$}{$\{ {}^ia_{t} - {}^ia_{t-1}, ({}^ib_{t}-{}^ib_{t-1}) / e^{{}^ia_{t-1}} \}$}
		\State \Return{ $\{{}^{t-1}a_{t}, {}^{t-1}b_{t}\}$ }
		\EndFunction
		
		\Function{AdjustForDrift}{$ {}^{t-1}a_t, {}^{t-1}b_t $}
		\Let{${}^{t-1}c_t$}{$ e^{{}^{t-1}a_t} + {}^{t-1}b_t $}
		\Let{$\delta$}{$(1-e^{{}^{t-1}a_t}) \times \xi_\text{gap}$}
		\Let{${}^{t-1}c_t$}{${}^{t-1}c_t - ({}^{t-1}c_t-1)\times \xi_\text{base} + \delta$}
		\Let{${}^{t-1}b_t$}{${}^{t-1}b_t - ({}^{t-1}b_t)\times \xi_\text{base} - \delta$}
		\State \Return{$ \{ \log({}^{t-1}c_t - {}^{t-1}b_t), {}^{t-1}b_t \} $}
		\EndFunction
	\end{algorithmic}
\end{algorithm}

\subsection{Estimating Spatial Parameters}
\label{estimating-spatial}
We observe that the temporal parameters $P'_t$ as estimated in~\ref{estimating-temporal} are robust enough to be considered as known while estimating $\mathcal{P}_s$. Hence, with known estimates for $P'_t$, similar to~\autoref{eq:g'functions}, using \autoref{eq:ghfunctions}, \autoref{eq:rela} and \autoref{eq:relb}, but without ignoring $h_{P_t,\mathcal{P}_s}$, we get a linear function  $f'_{\mathcal{P}_s}$ on $\mathcal{P}_s$
\begin{equation*}
\label{eq:f'functions}
I'_{n,t+1} = (I'_{m,t} - {}^tb_{t+1} + {}^pr_n-{}^pr_m ) / e^{{}^ta_{t+1}} = f'_{\mathcal{P}_s}(I'_{m,t})
\end{equation*}
Using pixel correspondences $C_t$, $P'_t$, and for $O_\text{TS}(f'_{\mathcal{P}_s}, C_t)$ in \autoref{eq:tsobjective} we can form a linear system of equations $A x = B$, where $x = \bar{\mathcal{P}}_s = [{}^pr_1, {}^pr_2, {}^pr_3, \ldots]^T$. Here $\bar{\mathcal{P}}_s \subset \mathcal{P}_s$ is the subset for which we have correspondences. Note that the $A$ matrix generated here would have a specific sparsity pattern - only two column blocks that correspond to residuals involved in the pixel correspondences will be nonzero. Thus, we can exploit this pattern to efficiently solve the sparse linear system. We can reduce the number of unknown variables from $|\Omega|$ by discretizing the image space uniformly or adaptively based on quadcode data structure~\cite{li1987quadcode}.

Note that a row in $A$ only ``connects'' two variables in $\mathcal{P}_s$. Even if matrix $A$ is full-rank, disconnected components in $\mathcal{P}_s$ can have unknown offsets as we recall that these parameters are w.r.t.\ a hypothetical frame without residual errors. One may choose to connect these disconnected components using heuristic, however, we simply choose to consider the largest connected component and hope that the connection grows over time as new pixel correspondences are added. To estimate the set of parameters $\mathcal{P}_s \setminus \bar{\mathcal{P}}_s$ for the pixels where we did not observe a correspondence, we employ a Gaussian Process Regression model \cite{kuss2006gaussian} trained on $\bar{\mathcal{P}}_s$, due to its non-linear properties and high expressivity.

\subsection{Photometric Calibration}
\label{photometric-calibration}
A calibrated pixel intensity for any 3D object viewed at pixel $m \in \Omega$ in any frame $t$ should be same as its intensity when viewed at pixel $p \in \Omega$ in the hypothetical frame $o$. Thus, an uncalibrated intensity $I'_{m,t}$, is calibrated using \autoref{eq:calib_model} as $I'_{p,o} = I'_{m,t}e^{{}^oa_t}+{}^ob_t-{}^pr_m$ (similarly each pixel in the image can be calibrated parallelly). The hypothetical frame $o$ is unobservable, hence we assume ($e^{{}^oa_1} = 1$, ${}^ob_1=0$), and use (\autoref{eq:rela}, \autoref{eq:relb}) to compute ($e^{{}^oa_t}, {}^ob_t$) from ($e^{{}^1a_t}, {}^1b_t$) respectively as estimated in~\ref{estimating-temporal}.

The above assumption, however might lead to unnatural intensity values $I'_{p,o} > 1$ or $I'_{p,o} < 0$. We adjust the unnatural values using a cyclic colormap $\bar{I}'_{p,o} = \textsc{Colormap}(I'_{p,o} \bmod 1.0)$. We want to choose a cyclic colormap which has high contrast and preferably is perceptually uniform. Experimentally, we found D3 Cyclic Rainbow~\cite{petzoldt2010cyclicrainbow} to be a good fit. However, if the downstream pipeline requires Grayscale images, it is better to use a cyclic grayscale ramp.
\begin{equation*}
\label{eq:grayscale_ramp}
\bar{I}'_{p,o} = \begin{cases}
2 (I'_{p,o} \bmod 1.0) &\text{if $(I'_{p,o} \bmod 1.0) < 0.5$,}
\\
-2 (I'_{p,o} \bmod 1.0)+2 &\text{otherwise}.
\end{cases}
\end{equation*}

\subsection{Computing Pixel Correspondences}
\label{sec:pixel_correspondences}
Any method can be used to compute pixel correspondences between two frames \cite{suhr2009kanade,sarlin2020superglue}. Note that this step depends on pixel appearances and thus is prone to photometric errors. Hence, there exists a Chicken-and-Egg problem. We are attempting to correct for photometric error using correspondences that are affected by such errors. Due to these errors, correspondence search between any two frames would suffer. However in such a scenario, we can provide correspondences from many older frames. Note that only two good correspondences are sufficient to estimate $P'_{i,t}$ between two frames, and $\mathcal{P}_s$ which needs to be estimated only once, can be estimated after accumulating many correspondences.

\section{Experiments and Results}
\label{results}
We analyze the performance of our proposed method for photometric calibration of thermal cameras through extensive qualitative and quantitative experiments, and validate the design choices made using ablation studies. Recall that the motivation for this work comes from the observation that TIR cameras exhibit high amount of 1) temporal, 2) spatial photometric variations, and have 3) low contrast which makes any downstream robotics perception pipeline primarily developed for RGB cameras prone to failure. 

For our application, we experimented with popular monocular VO/VIO algorithms (DSO \cite{engel2018direct}, SVO2.0 \cite{forster2016svo}, VINS-MONO \cite{qin2018vins}, MSCKF \cite{mourikis2007multi}) to localize our MAV at night \cite{daftry2018robust}. While these pipelines vary widely in their approach, there exists a common point of failure which can be contributed to the failure to track features from one frame to another under the three characteristic challenges as described above. Thus in \ref{highlevel}, we present the results related to only one such pipeline (SVO2.0).

\textbf{Baseline:} To the best of our knowledge, we did not find an algorithm for online photometric calibration of TIR cameras. However \cite{bergmann2018online} presents such an algorithm for RGB cameras, and hence can be a natural first choice for many. Using it as a baseline also enables us to study its performance for TIR.

Since our method can potentially benefit a wide range of computer vision pipelines that sit between raw camera images and downstream applications, we study the benefits of improved photometric calibration in two scenarios: 1) low-level vision tasks like feature tracking that forms the foundation for many CV applications, and 2) down-stream CV pipelines like VO and SLAM that are critical to enabling vison-based autonomous navigation.

\textbf{Experimental Setup:}
\label{setup}
We evaluate our method on two datasets: First, an open-source dataset from KAIST \cite{choi2015all} that consists of $50$km long video sequences in urban environment from a vehicle-mounted TIR camera, captured across different illumination conditions (time of day; e.g. AM02 represent data captured at 2am). Second, a custom dataset (named as MAV\_*) captured using a MAV mounted FLIR A65, a popular uncooled $\mu$BA TIR camera, in an outdoor natural environment with varied viewing angles. Both the dataset contains raw grayscale sequential images in standard \textit{uint8} encoding. No other camera parameters were used from the camera driver. The KAIST dataset exhibits \textbf{high temporal variations}, while the MAV dataset exhibits \textbf{high spatial variations}. We ran experiments for three versions of each dataset, namely uncalibrated (\textbf{Raw}), calibrated using baseline (\textbf{Baseline}), and calibrated using the proposed method (\textbf{Our}). In our implementation we tried KLT \cite{suhr2009kanade}, SuperGlue \cite{sarlin2020superglue}, and a feature tracker based on the front-end of \cite{engel2018direct} to generate correspondences. We observed that the choice depends on the dataset and computational budget. Except for \cite{sarlin2020superglue} which utilizes a GPU, the rest can run at $>60$fps on a portable CPU (TX2).

\subsection{Benefits to Feature Tracking}
\label{low-level-tracking}
Broadly, there exists three types of tracking: direct methods that uses raw image intensities to track features, indirect methods that compute feature descriptors, and rely on descriptor matching to keep track of features, and semi-direct methods that uses combination of direct and indirect methods. Here, we study only direct and semi-direct methods as they have been shown to be more suitable for thermal-images \cite{khattak2019keyframe}. The quality of feature tracking depends primarily on: 1) number of good features found in a frame (affected by the image contrast), 2) number of good features tracked from one frame to another (affected by temporal variations), and 3) number of frames a feature was successfully tracked through, also defined as feature-persistence (affected by both temporal and spatial variations).

\begin{table}[t!]
	\begin{center}
		\caption{Performance of KLT and SVO 2.0}
		\label{table:klt_svo}
	\begin{tabular}{ |c|c|c|c||c|c|c| } 
		\hline
		& \multicolumn{3}{|l||}{Mean number of frames} & \multicolumn{3}{|l|}{Mean features per frame}\\
		& \multicolumn{3}{|l||}{a KLT feature persisted } & \multicolumn{3}{|l|}{tracked by SVO 2.0}\\
		\hline
		\textbf{Dataset} & \textbf{Raw} & \textbf{Baseline} & \textbf{Our} & \textbf{Raw} & \textbf{Baseline} & \textbf{Our}\\
		\hline
		AM02 & 95 & 88 & \textbf{110} & 79 & 66 & \textbf{91}\\
		AM05 & 67 & 61 & \textbf{68} & 98 & 79 & \textbf{114}\\
		AM09 & 90 & 79 & \textbf{104} & 87 & 30 & \textbf{107}\\
		MAV\_A & 56 & 51 & \textbf{59} & 182 & 120 & \textbf{181}\\
		MAV\_B & 18 & 15 & \textbf{20} & 174 & 93 & \textbf{176}\\
		MAV\_C & \textbf{218} & 199 & 193 & 173 & \textbf{180} & 150\\
		\hline
	\end{tabular}
\end{center}
\vspace*{-1em}
\end{table}
\begin{table}[t!]
	\begin{center}
		\caption{Average number of features tracked through worst-events and percentage improvement w.r.t.\ Raw}
		\label{table:recovery}
		\begin{tabular}{ |c|c|c|c| } 
			\hline
			\textbf{Dataset} & \textbf{Raw} & \textbf{Baseline} & \textbf{Our} \\
			\hline
			AM02 & 299 & 921 (208 \%) & \textbf{1110 (271 \%)}\\
			AM05 & 319 & 481 (51 \%) & \textbf{609 (91 \%)}\\
			AM09 & 63 & 561 (790 \%) & \textbf{642 (919 \%)}\\
			MAV\_A & 223 & 216 (-3 \%) & \textbf{312 (40 \%)}\\
			MAV\_B & 90 & 63 (-30 \%) & \textbf{157 (74 \%)}\\
			MAV\_C & \textbf{830} & 767 (-8 \%) & 784 (-6 \%)\\
			\hline
		\end{tabular}
	\end{center}
	\vspace*{-3em}
\end{table}

For direct methods, we evaluate our method for tracking performance in the popular Kanade-Lucas-Tomasi (KLT) Feature Tracking \cite{suhr2009kanade} algorithm as this algorithm and it's derived versions forms a basis for many CV front-ends. The average feature-persistence for each dataset is presented in \autoref{table:klt_svo}. Loss of a feature may occur primarily when the feature goes out of the frame or KLT search window, and when KLT fails to track due to photometric error. For a dataset the affect of the former is similar across the three dataset versions, and only the later is the primary source of discrepancy. Except for the ``MAV\_C'' dataset, while the proposed method performs the best, surprisingly, the baseline performs worse than the raw. Recall that temporal parameters $P_t$ drifts over time by accumulating error (\ref{estimating-temporal}). When they drift outside the allowed domain, they lead to unnatural intensity values $I'_{p,o} > 1$ or $I'_{p,o} < 0$. The baseline solves this issue by artificially adjusting the exposure parameters to progress towards its nominal values similar to our method \textsc{AdjustForDrift}, but more aggressively and is susceptible to vanishing contrast in the image. The aggressive correction leads to drastic photometric change and reduction in contrast. This happens more frequently in TIR camera as compared to RGB cameras. Thus, while the baseline did perform photometric calibrations, however it eventually lead to more photometric errors everytime it had to correct for the frequent drifts observed in TIR cameras. In the proposed method we correct for this drift by softly tugging the parameters towards its nominal values but accounting for the case when they are outside their domain by employing a cyclic colormap (\ref{photometric-calibration}). This in addition also helps us maintain high image contrast.

\begin{figure}[t!]
    \vspace*{-1em}
	\centering
	\subfloat{\includegraphics[width=0.5\linewidth]{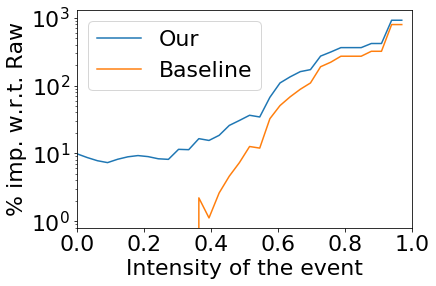}}		\subfloat{\includegraphics[width=0.5\linewidth]{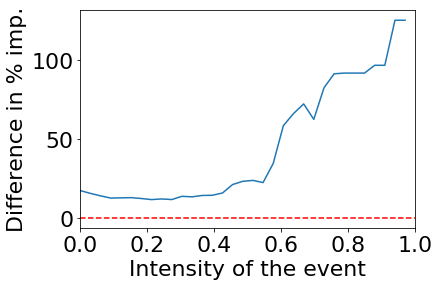}}
	\caption{\small \textbf{Left}: The log-graph shows that on average our method works better than the baseline for events of all magnitudes. \textbf{Right}: The difference in \% imp.\ between our method and baseline increases drastically as the magnitude increases. Worst-event at 1.}
	\label{fig:all_events}
	\vspace*{-2em}
\end{figure}

While small temporal variations are very common, however robotic applications are more concerned about large variations which can lead to failure even when they are rare. The goal of photometric calibration is to reduce these large variations in the data such that down-stream applications are less prone to failure. We perform a worst-case analysis on feature-tracking for each dataset. Let us define a worst-event as the occurrence of a large temporal variation similar to the worst temporal variation observed in the dataset. \autoref{table:recovery} presents the average number of features that were successfully tracked during such worst-events. The more the features, the less prone is the down-stream application to failure. We observe significant improvement of the proposed method over both the baseline and raw datasets. For the ``MAV\_*'' datasets, we observe a negative improvement for the baseline method. To understand the reason, recall that the ``MAV\_*'' datasets contain large spatial variations. The baseline method is developed for RGB cameras and thus assumes a radial model for spatial variations which is not suitable for TIR cameras. This, modelling error leads to worse than uncalibrated performance in these datasets. Next in \autoref{fig:all_events}, we also present performance for all magnitude of events (averaged over all the datasets) and not only the worst events. The magnitudes measure the amount of temporal variation and are presented here after normalization with that of the worst event.

Finally, for semi-direct methods, we use the front-end of the popular visual odometry algorithm SVO2.0 \cite{forster2016svo}. Note that, SVO2.0 does account for illumination gain and offset: parameters related to temporal variation, yet we still observe significant improvement of performance after the proposed calibration (\autoref{table:klt_svo}). The performance of baseline can similarly be explained as that of direct method.

\subsection{Benefit to Visual Odometry}
\label{highlevel}
We evaluate the benefits of photometric calibration towards monocular VO algorithms, using SVO2.0~\cite{forster2016svo} as a representative. SVO2.0 failed to run reliably on these challenging datasets without calibration, thus, in this context we study and advocate robustness as the primary metric for evaluation rather than VO localization accuracy. 

\begin{figure}[t!]
	\centering
	\includegraphics[width=0.5\columnwidth]{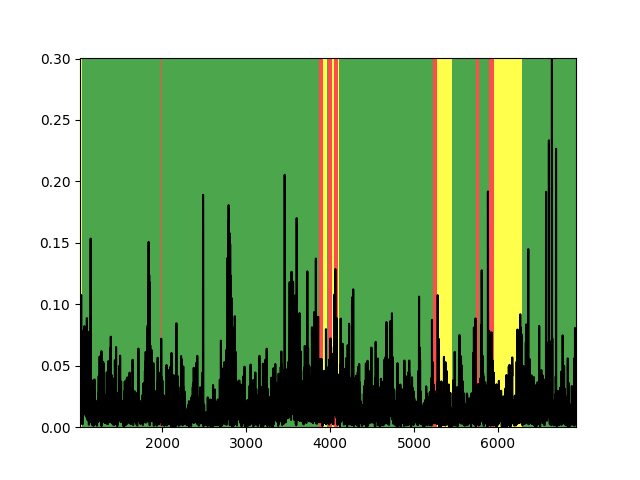}\hfill%
	\includegraphics[width=0.5\columnwidth]{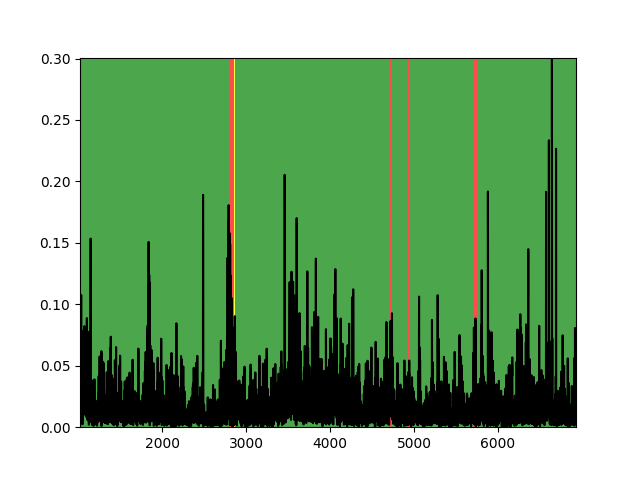}
	\caption{\small SVO2.0 Status over time. Black: measure of photometric error in each frame, Green: Good Tracking, Red: Bad Tracking, Yellow: Trying to initialize. Left: Uncalibrated. \textbf{Right}: Calibrated}
	\vspace{-5mm}
	\label{fig:svo-status}
\end{figure}

\autoref{table:svo-robustness} presents the improvement in robustness achieved as measured by two metrics: 1) The percentage of frames that are in ``Bad Tracking'', an 2) The number of times SVO2.0 failed completely and had to be restarted in the dataset. We define ``Bad Tracking'' as the status when SVO2.0 is less confident about the localization estimate. SVO2.0 publishes a ``Critical'' state message every time that happens (\autoref{fig:svo-status}). Note that SVO2.0 is a complete VO pipeline with its own robustness measures that include semi-direct feature tracking, keyframe based recovery, feature outlier rejection based on geometric constraints etc., yet we still observe high discrepancy in performance between ``Raw'' and ``Our'' versions. The proposed method clearly improves robustness significantly as there are fewer failures. We still observe some failures which might be due to general failure of SVO2.0 due to motion blur and low contrast of TIR images. However, surprisingly we observe high failure rate and ``Bad Tracking'' for the baseline method. This can again be contributed to the aggressive dift-correction, reduction of contrast and failure to model the spatial variation of TIR camera using a radial model as described in \ref{low-level-tracking}.

\setlength{\tabcolsep}{4pt}
\begin{table}[t!]
	\begin{center}
		\caption{Improvement in Robustness of SVO2.0}
		\label{table:svo-robustness}
		\begin{tabular*}{\linewidth}{l | @{\extracolsep{\fill}}  cccccc}
			\hline\noalign{\smallskip}
			Dataset & \multicolumn{3}{c}{Percentage of frames}  & \multicolumn{3}{c}{Num Failures}\\
			        & \multicolumn{3}{c}{in Bad Tracking}  & \\
			\noalign{\smallskip}
			\cline{2-4} \cline{5-7}
			\noalign{\smallskip}
			& Raw & Baseline & Our & Raw & Baseline & Our\\			
			\noalign{\smallskip}
			\hline
			\noalign{\smallskip}
			AM02 & 5.43 & 9.33 & \textbf{2.40} & 5 & 9 & \textbf{2}\\
			AM05 & 1.46 & 9.93 & \textbf{1.46} & 1 & 6 & 1\\
			AM09 & 3.16 & 5.75 & \textbf{0.0} & 2 & 4 & \textbf{0}\\
			MAV\_A & 0 & 12.32 & 0 & 0 & 1 & 0\\
			MAV\_B & 0 & 4.72 & 0 & 0 & 1 & 0\\
			\hline
		\end{tabular*}
		\vspace{-1mm}
	\end{center}
\end{table}
\setlength{\tabcolsep}{1.4pt}
\setlength{\tabcolsep}{4pt}
\begin{table}[t!]
	\begin{center}
		\caption{Mean \% of Photometric Error per pixel used to estimate $\mathcal{P}_s$ (Training) and the rest of the dataset (Validation)}
		\label{table:generalizability}
		\begin{tabular*}{\linewidth}{l | @{\extracolsep{\fill}}  cccccc}
			\hline\noalign{\smallskip}
			Dataset & \multicolumn{3}{c}{Training Data}  & \multicolumn{3}{c}{Validation Data}\\
			\noalign{\smallskip}
			\cline{2-4} \cline{5-7}
			\noalign{\smallskip}
			& w/o & with & with & w/o & with & with\\
			& Calib & $P'_t$ & $P'_t$ \& $\mathcal{P}_s$ & Calib & $P'_t$ & $P'_t$ \& $\mathcal{P}_s$\\			
			\noalign{\smallskip}
			\hline
			\noalign{\smallskip}
			AM02 & 4.88 & \textbf{1.06} & 1.24 & 5.20 & \textbf{1.30} & 1.50\\
			AM05 & 4.06 & \textbf{1.11} & 1.97 & 4.18 & \textbf{0.76} & 1.10\\
			AM09 & 4.30 & \textbf{0.89} & 1.30 & 4.42 & \textbf{0.84} & 1.40\\
			MAV\_A & 4.33 & 2.36 & \textbf{2.29} & 3.86 & 1.92 & \textbf{1.88}\\
			MAV\_B & 3.07 & 2.12 & \textbf{1.88} & 2.82 & 2.48 & \textbf{2.24}\\
			MAV\_C & 2.74 & 2.56 & \textbf{2.32} & 3.23 & 2.69 & \textbf{2.22}\\
			\hline
		\end{tabular*}
		\vspace{-3mm}
	\end{center}
\end{table}
\setlength{\tabcolsep}{1.4pt}

\setlength{\tabcolsep}{4pt}
\begin{table}[t!]
	\begin{center}
		\caption{Effect of Calibration with and without Drift adjustment. A higher coefficient means that the improvement is more evident for frames with higher photometric error.}
		\label{table:drift}
		\begin{tabular*}{\linewidth}{l | @{\extracolsep{\fill}}  cccc}
			\hline\noalign{\smallskip}
			Dataset & \multicolumn{2}{c}{$\xi_\text{base} = 0$}  & \multicolumn{2}{c}{$\xi_\text{base} = 0.025$}\\
			\noalign{\smallskip}
			\cline{2-3} \cline{4-5}
			\noalign{\smallskip}
			& $\rho$ & $\rho_l$, $\rho_u$ & $\rho$ & $\rho_l$, $\rho_u$\\		
			\noalign{\smallskip}
			\hline
			\noalign{\smallskip}
			AM02 & 0.14 & 0.03, 0.29 & \textbf{0.52} & \textbf{0.42, 0.62}\\
			AM05 & 0.33 & 0.15, 0.50 & \textbf{0.42} & \textbf{0.27, 0.55}\\
			AM09 & \textbf{0.56} & \textbf{0.47, 0.64} & 0.50 & 0.39, 0.59\\
			MAV\_A & -0.01 & -0.29, 0.27 & \textbf{0.39} & \textbf{0.10, 0.62}\\
			\hline
		\end{tabular*}
		\vspace{-3mm}
	\end{center}
	\vspace*{-1em}
\end{table}
\setlength{\tabcolsep}{1.4pt}

\subsection{Ablation Study}
\subsubsection{Generalization of Spatial Parameters}
We verify that the the GP Regression model do not overfit and generalize well. To this end, we estimate the average percentage of photometric error per pixel: first without any calibration (w/o Calib), then after applying only temporal calibration (with $P'_t$) and finally after applying both temporal and spatial calibration (with $P'_t$ \& $\mathcal{P}_s$). Training Data refers to $\bar{\mathcal{P}}_s$, while Validation Data refers to the set $\mathcal{P}_s \setminus \bar{\mathcal{P}}_s$ as defined in \ref{estimating-spatial}. As expected in \autoref{table:generalizability}, we observe that applying $\mathcal{P}_s$ reduces photometric error in MAV\_* dataset which exhibits significant spatial variations. And we observe that the photometric errors for Training and Validation Data are similar which indicates that the estimates do generalize well. This enables the method to compute $\bar{\mathcal{P}}_s \subset \mathcal{P}_s$ from sparse correspondences and trust the GP to predict the rest.

\subsubsection{With and Without Drift Adjustment}
Here we evaluate the contribution of Drift Adjustment and specifically letting $\xi_\text{base} > 0$. As described in \ref{estimating-temporal}, drift adjustment contains two components. One which attempts to maintain gap between the $c_t$ and $b_t$ parameters such that they do not collapse and to maintain contrast. And the other component attempts to pull them near their nominal values $c_t=1, b_t=0$. The former is crucial, and thus always required. However, we may ignore the latter and let the cyclic colormap handle the drift. This is applicable to scenarios when we do not want to introduce a bias on the parameters. However, it does affect low-level performance. \autoref{table:drift} presents the pearson correlation coefficient that relates 1) improvement in feature count between calibrated and uncalibrated data, and 2) magnitute of photometric error. The pearson coefficient $\rho$ is accompanied by its $95\%$ confidence interval $[\rho_l, \rho_u]$.

\subsubsection{Qualitative Analysis and Computation Time}
\begin{figure*}[t!]
	\centering
	\subfloat{\includegraphics[width=0.15\textwidth]{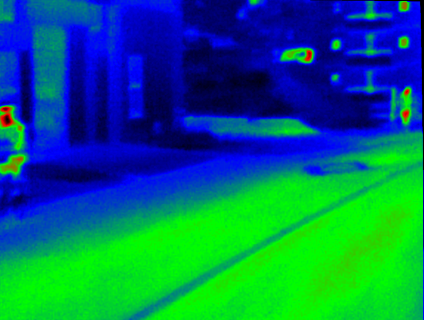}}
	\hfill
	\subfloat{\includegraphics[width=0.15\textwidth]{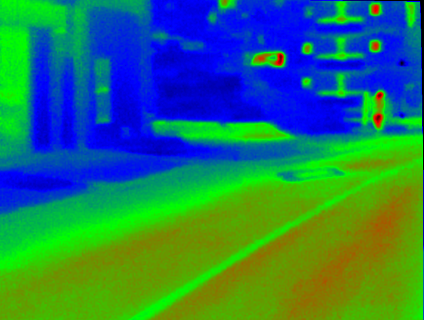}}
	\hfill
	\subfloat{\includegraphics[width=0.15\textwidth]{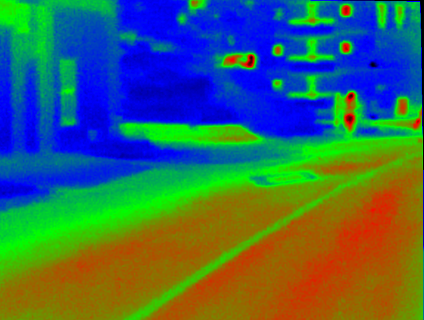}}
	\hfill
	\hfill
	\subfloat{\includegraphics[width=0.24\textwidth]{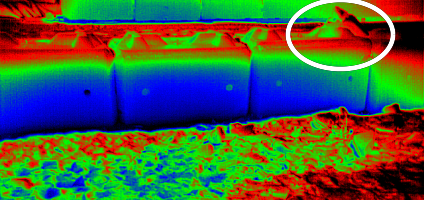}}
	\hfill
	\subfloat{\includegraphics[width=0.24\textwidth]{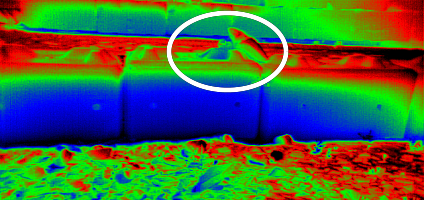}}
	\vspace*{-0.2cm}
	\addtocounter{subfigure}{-6}
	\subfloat{\includegraphics[width=0.15\textwidth]{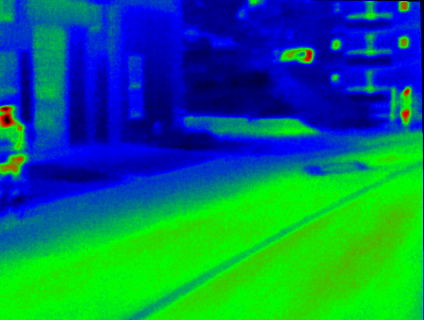}}
	\hfill
	\subfloat{\includegraphics[width=0.15\textwidth]{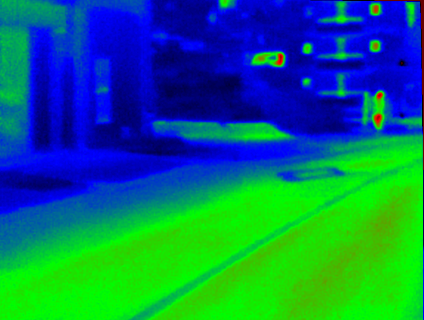}}
	\hfill
	\subfloat{\includegraphics[width=0.15\textwidth]{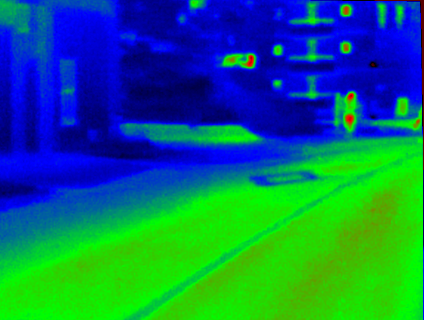}}
	\hfill
	\hfill
	\subfloat{\includegraphics[width=0.24\textwidth]{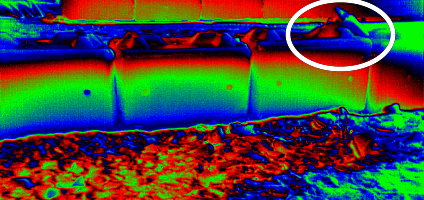}}
	\hfill
	\subfloat{\includegraphics[width=0.24\textwidth]{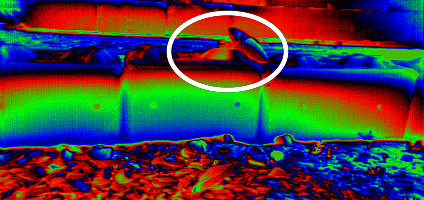}}	
	\caption{\small \textbf{Top:} Uncalibrated. \textbf{Bottom:} Calibrated. First three images show qualitative results on temporal calibration, while the last two shows spatial calibration. Notice how the appearance of an object (marked with white ellipse) changes when viewed at the top right corner of the image as compared to top center position. The camera towards the top corners appear to have accumulated spatial variations.}
	\label{fig:qualitative2}
	\vspace{-0.55cm}
\end{figure*}

For mapping, it is important that representation of a stationary 3D object as seen in some frame should not change in another frame. \autoref{fig:qualitative} and \autoref{fig:qualitative2} shows the qualitative results over few frames, and we observe that the consistency in representation would improve vision pipelines such as mapping.
The average computation time on an Intel Core i7-4710HQ CPU @ 2.50GHz is 0.01 second to estimate $P'_t$ per frame and 0.027 seconds to estimate $\mathcal{P}_s$ per iteration.

\section{Conclusion}
\label{conclusion}
In this paper we present a novel system for providing real-time online photometric calibration of thermal infrared cameras. Specifically, we model the sensor acquisition process, investigate the source of photometric error in such cameras and develop an efficient multi-threaded algorithm that can solve for the calibration parameters online. We study the benefits of our proposed method for both low- and high-level vision tasks, and demonstrate qualitative and quantitative improvements in performance over using uncalibrated thermal infrared images sequences. Finally, with this work we hope to take a step towards enabling widespread usage of vision-based methods for thermal-infrared cameras.

\section{Acknowledgments}
The research described in this paper was carried out at the Jet Propulsion Laboratory, Caltech, under a contract with the National Aeronautics and Space Administration.

\bibliographystyle{IEEEtran}
\bibliography{references}

\end{document}